\documentclass[aps,superscriptaddress,twocolumn]{revtex4-2}
\usepackage[utf8]{inputenc}
\usepackage{amsthm,amssymb,amsmath}
\usepackage{graphicx,xcolor}
\usepackage[colorlinks=true,linkcolor=blue,citecolor=blue,urlcolor=blue,linktocpage=true]{hyperref}

\usepackage{microtype}
\usepackage{booktabs}
\usepackage{float}
\usepackage{multirow}
 \usepackage{siunitx}
\usepackage[vlined,ruled]{algorithm2e}
\definecolor{cg}{rgb}{0,0.6,0}

\begin{document}
\title{Criticality-enhanced global frequency sensing with a monitored Kerr parametric oscillator via extended Kalman filter}

\author{Cheng Zhang}
\affiliation{Research Center for Quantum Physics and Technologies, Inner Mongolia University, Hohhot 010021, China}
\affiliation{School of Physical Science and Technology, Inner Mongolia University, Hohhot 010021, China}

\author{Mauro Cirio}
%\email{cirio.mauro@gmail.com}
\affiliation{Graduate School of China Academy of Engineering Physics, Haidian District, Beijing, 100193, China}

\author{Xin-Qi Li}
%\email{xinqi.li@imu.edu.cn}
\affiliation{Research Center for Quantum Physics and Technologies, Inner Mongolia University, Hohhot 010021, China}
\affiliation{School of Physical Science and Technology, Inner Mongolia University, Hohhot 010021, China}

\author{Pengfei Liang}
\email{pfliang@imu.edu.cn}
\affiliation{Research Center for Quantum Physics and Technologies, Inner Mongolia University, Hohhot 010021, China}
\affiliation{School of Physical Science and Technology, Inner Mongolia University, Hohhot 010021, China}

\date{\today}
\begin{abstract}
We analyze a global sensing scenario in which the frequency of a monitored Kerr parametric oscillator is estimated assuming limited prior information. The frequency is estimated in real-time by continuously monitoring the oscillator quadrature through homodyne detection and processing the resulting photocurrent with an extended Kalman filter (EKF). Due to the sensor nonlinearity, individual EKF trajectories do not always converge to the true unknown frequency in the long-time limit. However, we show that the statistical distribution of the frequency estimates does exhibit a sharp peak around the true value in the same limit. Leveraging this key statistical property, we develop a global sensing protocol assisted by adaptive control of the sensor parameters to harness critical enhancement. We present numerical evidence that this criticality-enhanced frequency estimation remains robust under low detection efficiency. 
\end{abstract}

\maketitle

\section{Introduction}

The emerging field of quantum metrology~\cite{Giovannetti2011,RevModPhys.89.035002,RevModPhys.90.035006,RevModPhys.90.035005,https://doi.org/10.1002/qute.202300218} exploits quantum resources to improve precision bounds. Quantum criticality~\cite{Sachdev_2011} is of particular interest in this context, owing to the extreme susceptibility of quantum states to external perturbations near quantum critical points. Within the scenario of local sensing, where small variation around a known parameter value is estimated, this property has been shown to enhance the precision bound from the standard quantum limit to the Heisenberg limit, or even the super-Heisenberg limit~\cite{PhysRevX.8.021022,PhysRevLett.124.120504,PhysRevLett.126.010502,DiCandia2023,PhysRevLett.132.060801,PhysRevLett.133.040801,PhysRevA.109.052604,Alushi2025,PhysRevLett.134.190802}. For example, a number of finite-component systems, such as the Rabi model~\cite{PhysRevLett.124.120504,PhysRevLett.126.010502,PhysRevLett.134.190802} and the Kerr parametric oscillator (KPO)~\cite{PhysRevLett.123.173601,DiCandia2023,PhysRevLett.133.040801,PhysRevA.109.052604,Alushi2025}, have been explored for achieving quantum-enhanced precision. On the other hand, global sensing~\cite{PhysRevLett.126.200501} of highly uncertain parameters presents additional challenges for critical metrology. These challenges can, in principle, be mitigated via feedback control~\cite{PhysRevLett.130.240803}. 

Recent studies~\cite{DiCandia2023,PhysRevLett.133.040801,PhysRevA.96.013817,PRXQuantum.3.010354,Montenegro2023,Ilias2024,gkdm-h2wx,PhysRevA.111.012412} have also extended the analysis of critical metrology from closed systems to Markovian open systems~\cite{10.1093/acprof:oso/9780199213900.001.0001,doi:10.1142/12402}. In this setting, quantum criticality manifests in the steady-state of a Lindblad master equation rather than in the ground state of the sensor Hamiltonian. This kind of quantum sensors opens new avenues for implementing sensing protocols based on time-continuous monitoring of their emission fields~\cite{Wiseman_Milburn_2009,ALBARELLI2024129260}, where measurement outcomes fluctuate in time due to the unavoidable quantum backaction on the sensors. 

Kalman filter~\cite{10.1115/1.3662552,10.1115/1.3658902} is a widely used  Bayesian estimation method for processing time-continuous measurement signals. Originally, it was designed for linear sensing models~\cite{10.1115/1.3662552,10.1115/1.3658902}, where temporal convergence of the filter can be rigorously proven. Variants have also been developed to account for nonlinear sensing models, such as EKF~\cite{Simon,Crassidis} and unscented Kalman filter~\cite{Julier1997NewEO,882463}, among others. Recently, Kalman filter and its variants have successfully been applied to atomic magnetometers for mitigating quantum noise~\cite{Amoros-Binefa_2021,k7nk-lrwd} and tracking time-varying signals~\cite{PhysRevA.102.063716,amorosbinefa2025trackingtimevaryingsignalsquantumenhanced}.

In this work, we investigate frequency estimation using an open KPO under continuous homodyne detection (HD) of its quadrature variable as the sensor. We focus on the case where prior knowledge of the oscillator frequency is limited, corresponding to the scenario of global sensing~\cite{PhysRevLett.126.200501,PhysRevLett.130.240803}. The frequency estimates are updated in real time via an EKF, which takes the measured photocurrent as its input.

We first analyze the properties of this EKF, in particular, focusing on its temporal convergence and its statistical behavior. Our analysis shows that while the EKF does not always ensure converging estimation in the long-time limit, it produces frequency estimates whose large-time distribution develops a prominent peak around the true frequency. Moreover, we identify an optimal homodyne phase that maximizes the sharpness of this peak. These findings motivate the development of a global sensing protocol assisted by adaptive control of sensor parameters for criticality-enhanced precision. We also present numerical results that demonstrate the effectiveness of this protocol, highlighting its robustness against detection inefficiency and confirming its validity for a moderate number of experimental repetitions.

This paper is organized as follows. In Sec.~\ref{sec:setup}, we describe the sensor model and its measurement dynamics. The challenges for global frequency estimation with this setup are also outlined. Sec.~\ref{sec:EKF} details the implementation of an EKF for the KPO sensor. In Sec.~\ref{sec:EKFtest}, we analyze the temporal convergence and the statistical properties of the resulting EKF estimates. In Sec.~\ref{sec:MCsampling}, we introduce a global sensing protocol that incorporates adaptive control of the KPO parameters, and assess its estimation performance. Conclusions are presented in Sec.~\ref{sec:conclusions}.

\section{Sensor model and dynamics}\label{sec:setup}

In this section, we define the open KPO sensor, describe its conditional dynamics under continuous HD, and outline the challenges involved in achieving global frequency estimation with this setup. The first two parts largely follow our earlier work in Ref.~\cite{zhang2025enhancinginformationretrievalquantumoptical}. 

This sensor is described by the Hamiltonian
\begin{equation}\label{eq:KerrHam}
H_\omega = \omega a^\dagger a + \frac{\epsilon}{2}(a^{\dagger2} + a^2) + \chi a^{\dagger2} a^2, 
\end{equation}
where $\omega$ is the oscillator frequency, $\epsilon$ the amplitude of the parametric driving and $\chi$ the Kerr nonlinearity. In this work, we focus on estimating the frequency $\omega$. This kind of Hamiltonian can also be engineered in other finite-component systems~\cite{PhysRevLett.92.073602,PhysRevLett.126.010502,PhysRevLett.134.190802,PhysRevLett.99.050402,PhysRevLett.115.180404,PhysRevLett.124.120504,PhysRevLett.124.040404}. 

We consider a leaky oscillator with damping rate $\kappa$, which constantly emits radiation quanta into its surrounding environment. The quadrature $x_\varphi = x\cos\varphi + p\sin\varphi $ of the output field is monitored via continuous HD, where $\varphi\in(-\pi,\pi]$ is the homodyne phase,  $x=(a+a^\dagger)/\sqrt2$ and $p=i(a^\dagger-a)/\sqrt2$ are canonical quadrature operators. Under this monitoring, the conditional state $\rho_c$ of the KPO evolves according to the stochastic master equation (SME)~\cite{Wiseman_Milburn_2009}
\begin{equation}\label{eq:sme}
d\rho_c = -i[H_\omega, \rho_c]dt + \kappa \mathcal{D}[a]\rho_cdt + \sqrt{\eta\kappa}\mathcal{H}[ae^{-i\varphi}]\rho_cdw,  
\end{equation}
with detection efficiency $\eta\in[0,1]$. The superoperators $\mathcal{D}[a]\rho_c = a\rho_c a^\dagger - \frac12 \{a^\dagger a,\rho_c\}$ and $\mathcal{H}[ae^{-i\varphi}]\rho_c = ae^{-i\varphi} \rho_c + \rho_c a^\dagger e^{i\varphi} - \langle ae^{-i\varphi} + ae^{i\varphi}\rangle_c \rho_c$, where $ \langle \cdot\rangle_c = \text{Tr}[\cdot\rho_c]$ represents the expectation over $\rho_c$, account for Markovian damping and measurement backaction, respectively. The recorded photocurrent at time $t$ reads 
\begin{equation}\label{eq:photocurrent}
dy_t = \sqrt{2\kappa\eta}\langle x_\varphi\rangle_c dt + dw, 
\end{equation}
where $dw$ is a Wiener increment satisfying $\mathbb{E}[dw]=0$ and $\mathbb{E}[dw^2]=dt$. 

Without continuous monitoring ($\eta=0$), the unconditional steady-state $\rho_\text{ss}$, i.e., the stationary solution to Eq.~(\ref{eq:sme}) for $\eta=0$, exhibits two different kinds of phase transitions in the scaling limit $\chi\to0$, depending on the sign of $\omega$~\cite{PhysRevLett.36.1135,Drummond_1980,PhysRevA.95.012128,PhysRevA.98.042118,PhysRevA.106.033707}. Specifically, for $\omega>0$ ($\omega<0$), the system undergoes a continuous (first-order) phase transition from the normal phase to the $\mathbb{Z}_2$ symmetry-broken phase. The phase boundary for the continuous phase transition is given by $\epsilon_c(\omega) = \sqrt{\omega^2+\kappa^2/4}$~\cite{DiCandia2023,PhysRevLett.133.040801}. 

The primary interest of this work is criticality-enhanced frequency estimation with an open KPO sensor. For consistency, our analysis is restricted to positive oscillator frequencies, $\omega > 0$. In addition, throughout this work the sensor operates in the normal phase ($\epsilon<\epsilon_c$) and in the regime of vanishing Kerr nonlinearity, i.e., $\chi\to0$. This regime can be realized experimentally in several platforms~\cite{10.1063/1.4937922}. In this regime, the conditional state $\rho_c$ preserves its Gaussian character for any Gaussian initial state. We introduce the variance $\sigma_o= 2(\langle o^2\rangle_c - \langle o\rangle_c^2)$ and the covariance $\sigma_{o_1o_2}= \langle\{o_1,o_2\}\rangle_c - 2\langle o_1\rangle_c\langle o_2\rangle_c$ for oscillator operators $o$ and $o_{1,2}$. The conditional state $\rho_c$ can then be fully characterized by its mean vector and covariance matrix 
\begin{equation}
\mathbf{r} = (X, P)^\intercal,~~~\boldsymbol\sigma = \begin{pmatrix}\sigma_x & \sigma_{xp} \\ \sigma_{xp} & \sigma_p\end{pmatrix}, 
\end{equation}
where $X = \langle x\rangle_c$ and $P = \langle p\rangle_c$.
Their dynamics are governed by the stochastic differential equations~\cite{zhang2025enhancinginformationretrievalquantumoptical,PhysRevA.95.012116}
\begin{subequations}\label{eq:rsigmaeqs}
\begin{align}
\displaystyle d\mathbf{r} &\displaystyle= \mathbf{A} \mathbf{r}dt + \sqrt{\frac{\eta\kappa}{2}}(\boldsymbol\sigma-\mathbf{I})\mathbf{B}d\mathbf{w}, \label{eq:rsigmaeqs1} \\
\displaystyle\frac{d\boldsymbol\sigma}{dt} &\displaystyle= \mathbf{A}\boldsymbol\sigma + \boldsymbol\sigma \mathbf{A}^\intercal + \mathbf{D} - \eta\kappa(\boldsymbol\sigma - \mathbf{I})\mathbf{B}(\boldsymbol\sigma - \mathbf{I}), \label{eq:rsigmaeqs2}
\end{align}
\end{subequations}
where the coefficient matrices are
\begin{equation}
\mathbf{A} = \begin{pmatrix}
-\kappa/2 & \omega-\epsilon \\
-\omega-\epsilon & -\kappa/2
\end{pmatrix},~\mathbf{B} = \begin{pmatrix}
\cos^2\varphi & \sin\varphi\cos\varphi \\
\sin\varphi\cos\varphi & \sin^2\varphi
\end{pmatrix},
\end{equation}
and $\mathbf{D}=\text{diag}(\kappa,\kappa)$, $\mathbf{I}$ is the $2\times2$ identity matrix,  and $d\mathbf{w} = (\cos\varphi dw, \sin\varphi dw)^\intercal$. 

Global sensing~\cite{PhysRevLett.126.200501} refers to the scenario where prior knowledge of the unknown true parameter is highly uncertain. For the frequency estimation considered here, we denote the unknown frequency by $\omega_\text{true}$ and quantify this initial uncertainty by assuming that $\omega_\text{true}$ lies within an interval $I_\text{prior} = (\omega_l, \omega_h)$ with $\omega_h > \omega_l \ge0$, i.e.,
\begin{equation}\label{eq:prior}
\omega_\text{true}\in I_\text{prior} = (\omega_l, \omega_h). 
\end{equation}
Global frequency estimation corresponds to the case of a broad prior interval satisfying $2\lvert(\omega_h-\omega_l)/(\omega_h+\omega_l)\rvert \sim \mathcal{O}(1)$. Here, this lower constraint $\omega_\text{true} > \omega_l$ is a direct consequence of the fact that the open KPO undergoes continuous phase transitions only for positive frequencies $\omega > 0$~\cite{PhysRevLett.36.1135,Drummond_1980,PhysRevA.95.012128,PhysRevA.98.042118,PhysRevA.106.033707}. Crucially, we will utilize this constraint in the adaptive protocol (Sec.~\ref{sec:MCsampling}) to select the initial driving amplitude, ensuring the sensor always operates in the normal phase.

Such limited prior information poses a typical challenge for global sensing in critical metrology. In the present setup, this challenge can be summarized as two main questions: (i) How can the driving amplitude $\epsilon$ be tuned towards the unknown critical value $\epsilon_c(\omega_\text{true})$ to achieve criticality-enhanced precision? (ii) How should the homodyne phase $\varphi$ be selected to optimize sensing performance? In the remainder of this paper, we address these questions when an EKF is employed to produce frequency estimates.

\begin{figure}%[t!]
\includegraphics[clip,width=8.5cm]{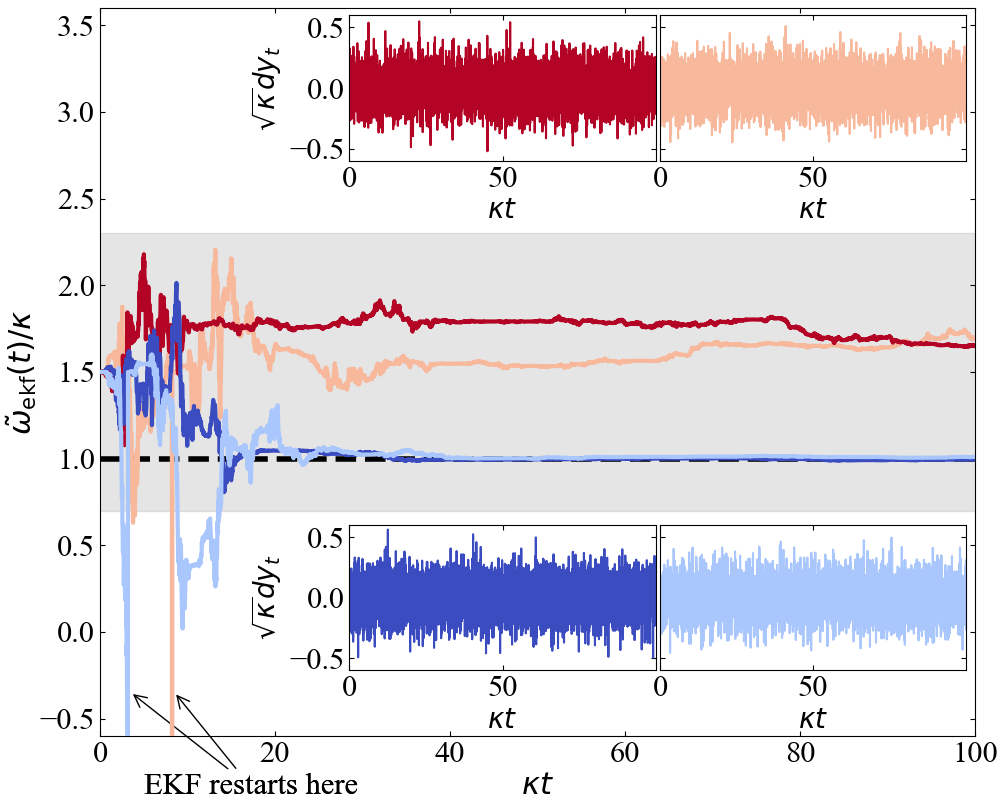}
\caption{Four typical trajectories of the EKF estimate $\tilde\omega_\text{ekf}(t)$ for ideal detection efficiency $\eta=1$ and $\varphi=0.1072$. The true frequency $\omega_\text{true}=\kappa$ is shown as the dashed line, and the prior interval $I_\text{prior} = (\omega_l, \omega_h)$, where $\omega_l=0.7\kappa$ and $\omega_h=2.3\kappa$, is displayed as the shaded gray area. The time instants where the filter divergence condition was triggered are marked. The insets show the corresponding photocurrents $\mathbf{Y}_{t}$, recorded with a time step $\kappa dt=0.02$. The EKF is initialized as $\tilde{\mathbf{x}}(0) = (0,0,1,1,0,(\omega_l+\omega_h)/2)^\intercal$ and $\tilde{\mathbf{\Sigma}}(0) = \text{diag}(10^{-3},10^{-3},10^{-3},10^{-3},10^{-3},\kappa^2v)$, with $v=1$ used here. The driving amplitude is $\epsilon=\kappa$ and the threshold parameter in Eq.~(\ref{eq:threshold}) is set to $F_\text{max}=10^5$.
}\label{fig:singletrajs}
\end{figure}

\section{Extend Kalman filter}\label{sec:EKF}

We begin by implementing an EKF for the KPO sensor. To this end, we first construct a state model based on Eq.~(\ref{eq:rsigmaeqs}). Specifically, the components of $\mathbf{r}$ and $\boldsymbol{\sigma}$ are combined into a state vector 
\begin{equation}
\mathbf{x}(t) = (X(t),P(t),\sigma_x(t),\sigma_p(t),\sigma_{xp}(t),\omega)^\intercal, 
\end{equation}
where the frequency $\omega$ is also included as the last component. The dynamical equations for $\mathbf{x}(t)$ and the photocurrent $dy_t$ are obtained by rewriting the differential equations in Eq.~(\ref{eq:rsigmaeqs}) and Eq.~(\ref{eq:photocurrent}) as 
\begin{subequations}\label{eq:stateeq}
\begin{align}
\displaystyle d\mathbf{x} &\displaystyle = \mathbf{F}_\mathbf{x} dt + \mathbf{G}_\mathbf{x}dw, \label{eq:stateeq1} \\
\displaystyle dy_t &\displaystyle= \mathbf{H}\mathbf{x}dt + dw, \label{eq:stateeq2}
\end{align}
\end{subequations}
where $\mathbf{F}_\mathbf{x}$ and $\mathbf{G}_\mathbf{x}$ are vector-valued functions mapping $\mathbf{x}$ to $6\times1$ column vectors, and $\mathbf{H}$ is a $1\times6$ row vector. Their explicit expressions are provided in Appendix~\ref{sec:sm_cm}. 

Importantly, the last term $- \eta\kappa(\boldsymbol\sigma - \mathbf{I})\mathbf{B}(\boldsymbol\sigma - \mathbf{I})$ in Eq.~(\ref{eq:rsigmaeqs2}) is nonlinear in $\boldsymbol\sigma$; consequently, $\mathbf{F}_\mathbf{x}$ is also a nonlinear function of $\mathbf{x}$, as seen in its third to fifth components in Eq.~(\ref{eq:Fx}). As will be discussed in Sec.~\ref{sec:EKFtest}, this nonlinearity has a significant impact on the performance of the EKF. 

\subsection{EKF equations}

In Bayesian inference~\cite{RevModPhys.90.035005}, estimates of unknown parameters are derived from a posterior distribution. In the context of the open KPO sensor considered here, this posterior is denoted as $P_\text{post}(\mathbf{z},t\vert \mathbf{Y}_{t})$, where the argument $\mathbf{z}$ is a real-valued $6\times1$ vector. For a specific photocurrent record $\mathbf{Y}_{t} = \{dy_\tau\vert 0\le \tau \le t\}$, the posterior distribution at time $t$ is updated via Bayes's rule 
\begin{equation}
P_\text{post}(\mathbf{z},t+dt\vert \mathbf{Y}_{t+dt}) = \frac{P_\text{sm}(dy_t\vert \mathbf{z})P_\text{post}(\mathbf{z},t\vert \mathbf{Y}_{t})}{\int d\mathbf{z}P_\text{sm}(dy_t\vert \mathbf{z})P_\text{post}(\mathbf{z},t\vert \mathbf{Y}_{t})}, 
\end{equation}
where $P_\text{sm}(dy_t\vert \mathbf{z})$ represents the conditional probability of obtaining the measurement outcome $dy_t$, provided that the state vector at time $t$ is $\mathbf{x}(t)=\mathbf{z}$. Because HD belongs to the class of Gaussian measurements, the conditional distribution $P_\text{sm}(dy_t\vert \mathbf{z})$ is Gaussian in terms of $dy_t$, with mean $\mathbf{H}\mathbf{z}dt$ and variance $dt$, as given by Eq.~(\ref{eq:stateeq2}). 

For nonlinear state models, such as the one defined in Eq.~(\ref{eq:stateeq}), the posterior generally becomes non-Gaussian even if a Gaussian posterior $P_\text{post}(\mathbf{z},t=0\vert \mathbf{Y}_{0}=\emptyset)$ at the initial time $t=0$ is assumed. The EKF addresses this difficulty by approximating the true posterior $P_\text{post}(\mathbf{z},t \vert \mathbf{Y}_{t})$ by a time-varying Gaussian distribution, which is completely characterized by a time-dependent $6\times1$ mean vector $\mathbf{\tilde{x}}(t)$ and a $6\times6$ covariance matrix $\mathbf{\tilde{\Sigma}}(t)$. 
We express $\mathbf{\tilde{x}}(t)$ explicitly as 
\begin{equation}
\mathbf{\tilde x}(t) = (\tilde{X}(t),\tilde{P}(t),\tilde\sigma_x(t),\tilde\sigma_p(t),\tilde\sigma_{xp}(t),\tilde\omega_\text{ekf}(t))^\intercal, 
\end{equation}
whose components are interpreted as the EKF estimates of the corresponding elements of the true state vector $\mathbf{x}(t)$ at time $t$. To retain the Gaussian character of the posterior, the nonlinear function $\mathbf{F}_\mathbf{x}$ should be expanded around the estimate $\mathbf{\tilde x}(t)$ when propagating the covariance $\mathbf{\tilde \Sigma}(t)$ at time $t$. Upon this expansion, the deterministic propagation of the covariance $\mathbf{\tilde \Sigma}(t)$ is governed by a $6\times6$ Jacobian matrix $\mathbf{F}'_{\mathbf{\tilde x}} = \nabla\mathbf{F}_{\mathbf{x}}\vert_{\mathbf{x} = \mathbf{\tilde x}}$. Explicit form of this Jacobian matrix is given in  Appendix~\ref{sec:sm_cm}.

Ultimately, the EKF propagates the mean $\mathbf{\tilde{x}}(t)$ and the covariance $\mathbf{\tilde \Sigma}(t)$ according to the following stochastic differential equations~\cite{k7nk-lrwd,amorosbinefa2025trackingtimevaryingsignalsquantumenhanced}
\begin{equation}\label{eq:ekfeqs}
\begin{array}{lll}
\displaystyle d\mathbf{\tilde x} &\displaystyle= \mathbf{F}_\mathbf{\tilde x} dt + \mathbf{K}_{\mathbf{\tilde x},\mathbf{\tilde\Sigma}}(dy_t - \mathbf{H}\mathbf{\tilde x}dt), \\
\displaystyle \frac{d\mathbf{\tilde\Sigma}}{dt} &\displaystyle= \left(\mathbf{F}'_\mathbf{\tilde x} - \mathbf{G}_\mathbf{\tilde x}\mathbf{H}\right)\mathbf{\tilde\Sigma} + \mathbf{\tilde\Sigma}\left( \mathbf{F}'_\mathbf{\tilde x} - \mathbf{G}_\mathbf{\tilde x}\mathbf{H} \right)^\intercal - \mathbf{\tilde\Sigma} \mathbf{H}^\intercal \mathbf{H} \mathbf{\tilde\Sigma},
\end{array}
\end{equation}
where 
$\mathbf{K}_{\mathbf{\tilde x},\mathbf{\tilde\Sigma}} = \mathbf{\tilde\Sigma}\mathbf{H}^\intercal + \mathbf{G}_\mathbf{\tilde x}$ represents the Kalman gain.

We emphasize that, owing to the nonlinearity of the state model~(\ref{eq:stateeq}), convergence of $\mathbf{\tilde{x}}(t)$ to the real state vector $\mathbf{x}(t)$ in the long-time limit $t\to+\infty$ is not guaranteed. As a consequence, the covariance matrix $\mathbf{\tilde \Sigma}(t)$ generally differs from the true error matrix $\mathbb{E}\left[\left(\mathbf{\tilde{x}}(t) - \mathbf{x}(t)\right)\left(\mathbf{\tilde{x}}(t) - \mathbf{x}(t)\right)^\intercal\right]$. 

\begin{figure}%[t!]
\includegraphics[clip,width=8.5cm]{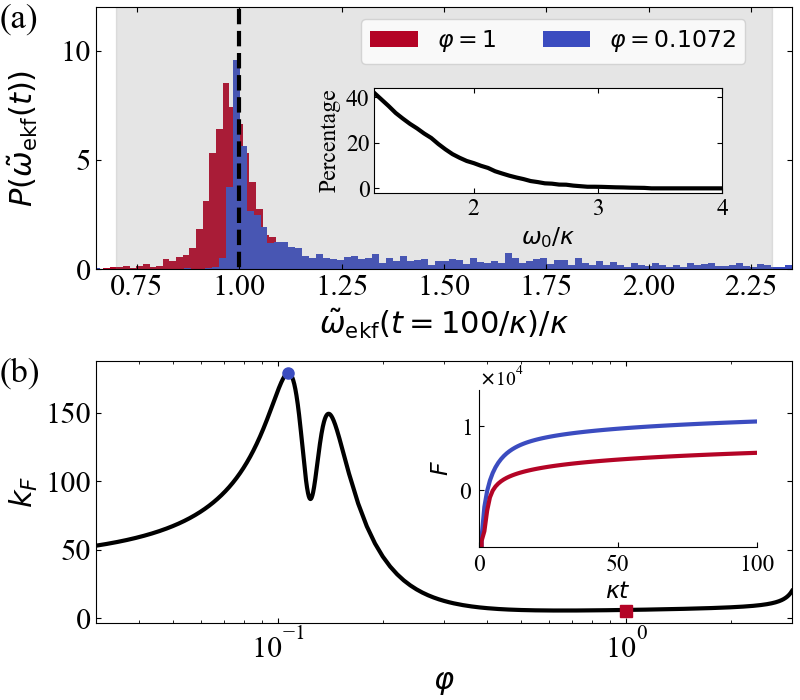}
\caption{(a) Distributions $P(\tilde{\omega}_\text{ekf}(t))$ at time $t=100/\kappa$ for two homodyne phases $\varphi=0.1072$ (blue) and $\varphi=1$ (red), obtained from $N_\text{traj}=2000$  trajectories under ideal HD ($\eta=1$). The vertical dashed line indicates the true frequency $\omega_\text{true}=\kappa$, and the gray shaded region corresponds to the prior interval $I_\text{prior}=(0.7\kappa,2.3\kappa)$. The inset shows the percentage of the $2000$ EKF estimates for $\varphi=0.1072$ that are larger than a certain frequency $\omega_0$. (b) Long-time growth rate $k_F$ of the CFI $F(t;\varphi,\omega,\epsilon,\eta)$ plotted as a function of the homodyne phase $\varphi$. The blue dot and red square mark the phase values $\varphi=0.1072$ and $\varphi=1$ used in panel (a), respectively. The inset shows the time evolution of the CFI $F$ at these two phase values. All other simulation parameters are the same as those in Fig.~\ref{fig:singletrajs}.
}\label{fig:dist}
\end{figure}

\subsection{Discretized EKF equations for numerical simulations}

The covariance matrix $\mathbf{\tilde \Sigma}(t)$ must remain positive definite to ensure a meaningful posterior distribution. Nevertheless, a straightforward Euler discretization of the stochastic differential equations in Eq.~(\ref{eq:ekfeqs}), i.e., replacing the differentials $dt$, $dy_t$ by their finite differences $\Delta t$, $\Delta y_t$ and retaining only terms up to the first order in the time step $\Delta t$, fails to preserve this property. Here, we present a discretization scheme capable of maintaining the positivity of $\mathbf{\tilde \Sigma}(t)$ throughout the filtering process. 

We divide the total measurement duration $T$ into time bins with length $\Delta t$, and denote the mean vector and the covariance matrix in Eq.~(\ref{eq:ekfeqs}) at time $t=n\Delta t$ as $\mathbf{\tilde x}_n = \mathbf{\tilde x}(t=n\Delta t)$ and $\mathbf{\tilde \Sigma}_n=\mathbf{\tilde \Sigma}(t=n\Delta t)$. The proposed discretization scheme proceeds in two steps:

\emph{1. Prediction:} This step accounts for the evolution driven by terms in Eq.~(\ref{eq:ekfeqs}) that are independent of the measurement outcome $dy_t$, such as $\mathbf{F}_\mathbf{\tilde x}$ and $\mathbf{F}'_\mathbf{\tilde x} - \mathbf{G}_\mathbf{\tilde x}\mathbf{H}$. We thus compute an intermediate vector $\mathbf{\bar x}_{n+1}$ and matrix $\mathbf{\bar \Sigma}_{n+1}$ via
\begin{equation}
\begin{array}{lll}
\displaystyle \mathbf{\bar x}_{n+1} &\displaystyle= \mathbf{\tilde x}_{n} + \mathbf{F}_{\mathbf{\tilde x}_{n}}\Delta t, \\
\displaystyle \mathbf{\bar \Sigma}_{n+1} &\displaystyle= \left[1 + \left(\mathbf{\tilde F}_{\mathbf{\tilde x}_{n}} - \mathbf{G}_{\mathbf{\tilde x}_{n}}\mathbf{H}\right)\Delta t\right]\mathbf{\tilde \Sigma}_{n}\\
&\displaystyle~~~\times\left[1 + \left(\mathbf{\tilde F}_{\mathbf{\tilde x}_{n}} - \mathbf{G}_{\mathbf{\tilde x}_{n}}\mathbf{H}\right)\Delta t\right]^\intercal. 
\end{array}
\end{equation}

\emph{2. Update:} The intermediate quantities are then updated using the remaining terms in Eq.~(\ref{eq:ekfeqs}) that depend on the measurement outcome: 
\begin{equation}
\begin{array}{lll}
\displaystyle \mathbf{\tilde x}_{n+1} &\displaystyle= \mathbf{\bar x}_{n+1} + \mathbf{K}_{\mathbf{\bar x}_{n+1},\mathbf{\bar\Sigma}_{n+1}}\left(\Delta y_t - \mathbf{H}\mathbf{\bar x}_{n+1}\Delta t\right), \\
\displaystyle \mathbf{\tilde \Sigma}_{n+1} &\displaystyle= \left(1 - \mathbf{\bar\Sigma}_{n+1}\mathbf{H}^\intercal \mathbf{H}\Delta t\right)\mathbf{\bar \Sigma}_{n+1}\left(1 - \mathbf{\bar\Sigma}_{n+1}\mathbf{H}^\intercal \mathbf{H}\Delta t \right)^\intercal\\
&\displaystyle~~~ + \mathbf{\bar\Sigma}_{n+1}\mathbf{H}^\intercal \mathbf{H}\mathbf{\bar\Sigma}_{n+1} \Delta t. 
\end{array}
\end{equation}
Clearly, the above equations for $\mathbf{\bar \Sigma}_{n+1}$ and $\mathbf{\tilde \Sigma}_{n+1}$ are both invariant under transposition, ensuring their positive definiteness throughout the iteration as long as the initial covariance $\mathbf{\tilde \Sigma}_0$ is set to be positive.

\subsection{Restarting the EKF in case of filter divergence}

The EKF equations~(\ref{eq:ekfeqs}) may produce a divergent estimate $\mathbf{\tilde{x}}(t)$ due to the potential dynamical instability in Eq.~(\ref{eq:rsigmaeqs}). Specifically, instability arises  when the estimated frequency $\tilde{\omega}_\text{ekf}(t)$ causes one eigenvalue of the matrix $\mathbf{\tilde A} = (-\kappa/2,\tilde{\omega}_\text{ekf}(t)-\epsilon;-\tilde{\omega}_\text{ekf}(t)-\epsilon,-\kappa/2)$ to become positive. In such cases, the Frobenius norm $\| \mathbf{\tilde{x}} \|_F = \sqrt{\tilde{X}^2 + \tilde{P}^2 + \tilde{\sigma}_x^2 + \tilde{\sigma}_p^2 + \tilde{\sigma}_{xp}^2 + \tilde{\omega}_\text{ekf}^2/\kappa^2}$ tends to grow exponentially within a few time steps, ultimately leading to a divergent EKF estimate. To mitigate this issue, we employ a restarting strategy: the Frobenius norm $\| \mathbf{\tilde{x}} \|_F$ is monitored continuously during the filtering. Once $\| \mathbf{\tilde{x}} \|_F$ exceeds a preset threshold $F_\text{max}$, i.e., 
\begin{equation}\label{eq:threshold}
\| \mathbf{\tilde{x}} \|_F > F_\text{max},
\end{equation}
the filter is re-initialized, and the estimation process restarts at that time instant. This strategy ensures that the estimation remains bounded throughout the entire filtering procedure.

\section{Properties of the EKF equations~(\ref{eq:ekfeqs})}\label{sec:EKFtest}

In this section, we present numerical results for analyzing two important properties of the EKF equations~(\ref{eq:ekfeqs}). Specifically, in Sec.~\ref{sec:tcong} we numerically examine the temporal convergence of individual trajectories of the EKF estimate $\tilde{\omega}_\text{ekf}(t)$, specifically, whether $\tilde{\omega}_\text{ekf}(t)$ approaches the unknown frequency $\omega_\text{true}$ as $t\to+\infty$. This helps to assess the feasibility of achieving single‑shot estimation via the EKF for the open KPO sensor. In Sec.~\ref{sec:PDFvphi}, we discuss the probability distribution of $\tilde{\omega}_\text{ekf}(t)$ at a fixed time $t$, in particular focusing on its dependence on the homodyne phase $\varphi$. 

We emphasize that the present analysis does not yet simulate a full global frequency estimation task. As noted at the end of Sec.~\ref{sec:setup}, such a task would require a dedicated strategy for controlling the two sensor parameters $\epsilon$ and $\varphi$. In fact, the results presented here provide the motivation for the adaptive sensing protocol developed in Sec.~\ref{sec:MCsampling}.

Throughout our numerical simulations in this work, the KPO sensor is always initialized in its vacuum state  Combining this condition with the prior knowledge represented by the interval $I_\text{prior}$, we initialize the EKF as $\tilde{\mathbf{x}}(0) = (0,0,1,1,0,(\omega_l+\omega_h)/2)^\intercal$ and $\tilde{\mathbf{\Sigma}}(0) = \text{diag}(10^{-3},10^{-3},10^{-3},10^{-3},10^{-3},v\kappa^2)$.  The initial frequency guess is set as $\tilde{\omega}_\text{ekf}(0)=(\omega_l+\omega_h)/2$, and the choices $\tilde{X}(0)=\tilde{P}(0)=0$, $\tilde{\Sigma}_x(0)=\tilde{\Sigma}_p(0)=1$ and $\tilde{\Sigma}_{xp}(0)=0$ are due to the initial vacuum state. The uncertainty $v\kappa^2$ of $\tilde{\omega}_\text{ekf}(0)$ is set on the order of $v\kappa^2\sim \lvert\omega_h - \omega_l\rvert^2$, while the small uncertainty $10^{-3}$ for the other EKF components is introduced solely to  avoid numerical singularities.

\subsection{Temporal convergence of individual trajectories}\label{sec:tcong}

To highlight different dynamical behaviors, in Fig.~\ref{fig:singletrajs}, we plot 
four typical trajectories of the EKF estimate $\tilde{\omega}_\text{ekf}(t)$ up to time $\kappa t=100$, assuming ideal detection efficiency ($\eta=1$). The corresponding photocurrent records $\mathbf{Y}_{t\ge0}$ supplied to the EKF are displayed in the insets. Impressively, the EKF is able to produce frequency estimates (e.g., the two blue curves) converging to the true frequency $\omega_\text{true}=\kappa$ (black dashed line) as $t \to +\infty$ for most photocurrent records, even though the records themselves appear extremely noisy.  However, there also exist a fraction of trajectories (e.g., the two red ones) yielding erroneous estimates. 

We emphasize that, this failure to converge cannot be attributed to the occurrence of filter divergence. Indeed, our numerical simulations reveal the presence of both converging trajectories that trigger the filter divergence condition~(\ref{eq:threshold}) (such as the light blue curve) and non‑converging ones that do not (such as the dark red curve). Rather, since this lack of convergence can occur irrespective of the time $t$, it must stem from the inherent nonlinearity of the state model~\eqref{eq:stateeq}, as noted at the beginning of Sec.~\ref{sec:EKF}.

These observations lead us to conclude that single-shot estimation, i.e., inferring the unknown frequency from a single trajectory, is generally not feasible for the open KPO sensor using the EKF alone. Nevertheless, we also emphasize that this does not preclude the possibility of achieving one-shot estimation by integrating the EKF with an appropriate real-time feedback control strategy. Indeed, such a combined approach has been successfully demonstrated in atomic magnetometers~\cite{k7nk-lrwd,amorosbinefa2025trackingtimevaryingsignalsquantumenhanced} for the real-time tracking of fluctuating magnetic fields~\cite{amorosbinefa2025trackingtimevaryingsignalsquantumenhanced}. We leave the exploration of this direction for future work.

\begin{figure*}%[t!]
\includegraphics[clip,width=0.9\textwidth]{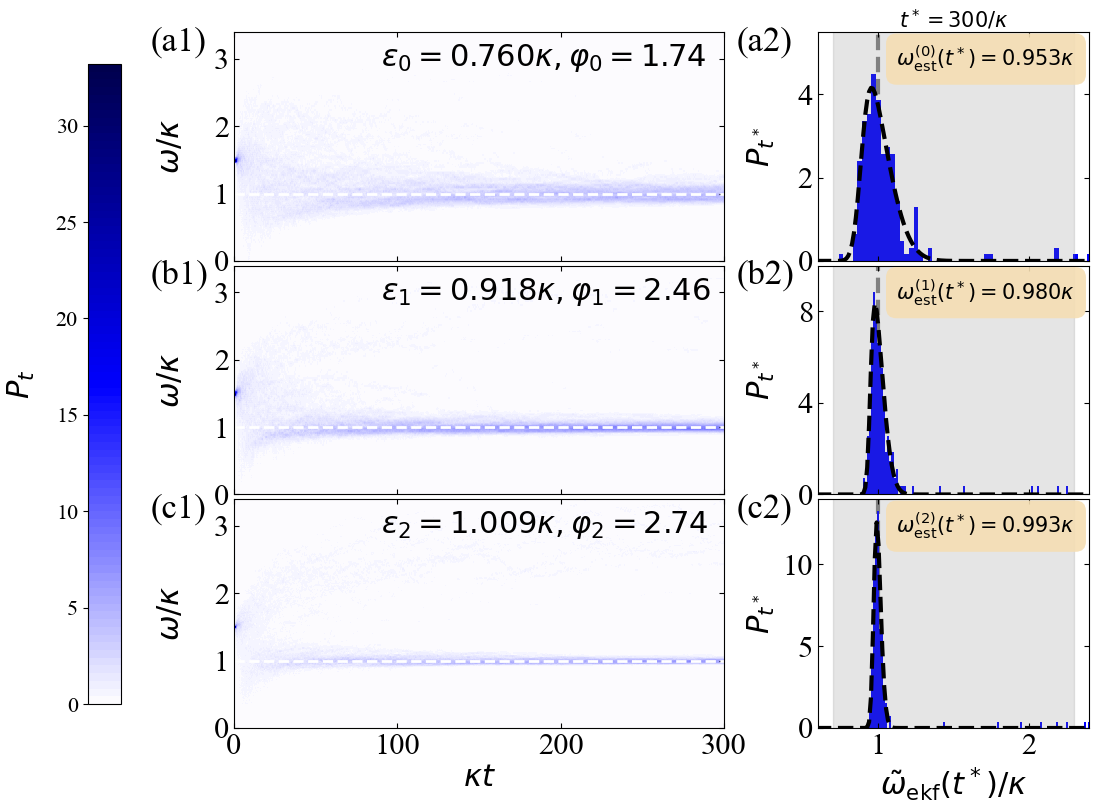}
\caption{(a1)-(c1) Time evolution of the distribution $P_t\equiv P(\tilde{\omega}_\text{ekf}(t))$ for the first three iterations ($i=0,1,2$) of the proposed protocol, obtained with $N_\text{traj}=200$ trajectories of $\tilde{\omega}_\text{ekf}(t)$ for a low detection efficiency $\eta=0.2$ and the prior interval $I_\text{prior} = (0.7\kappa, 2.3\kappa)$. The control parameters $(\epsilon_i, \varphi_i)$ are selected via the rules~(\ref{eq:selectrule}), using the update function $\mathcal{E}(\epsilon',\epsilon'') = (\epsilon' + \epsilon'')/2$. The true frequency is $\omega_\text{true}=\kappa$ (white dashed lines). (a2)-(c2) Distributions at time $t^*=300/\kappa$, together with their numerical fits (black dashed lines) to the skew‑normal function in Eq.~(\ref{eq:skn}). Vertical dashed lines mark $\omega_\text{true}=\kappa$ and gray shaded areas indicate the prior interval $I_\text{prior}$. 
}\label{fig:adaptiveEKF}
\end{figure*}

\subsection{Probability distribution of the EKF estimate and its relation to the classical Fisher information}\label{sec:PDFvphi}

Although the EKF estimate $\tilde{\omega}_\text{ekf}(t)$ does not ensure convergence to the unknown true frequency $\omega_\text{true}$ over time, its probability distribution $P(\tilde{\omega}_\text{ekf}(t))$ at a sufficiently large, fixed time $t$ exhibits a prominent peak centered around $\omega_\text{true}$. This is demonstrated in Fig.~\ref{fig:dist}(a), which shows two such distributions at time $t=100/\kappa$ obtained from $N_\text{traj}=2000$ trajectories under identical driving amplitude $\epsilon$ but with two different homodyne phases $\varphi=0.1072$ (blue) and $1$ (red). Interestingly, the distribution peak for $\varphi=0.1072$ is visibly sharper than that for $\varphi=1$. We now show that this important statistical property is directly linked to the long-time behavior of the classical Fisher information (CFI), here denoted explicitly as $F(t;\varphi,\omega,\epsilon,\eta)$ to highlight its dependence on all relevant parameters. 

In the Gaussian limit $\chi\to0$, the CFI can be computed via~\cite{PhysRevA.95.012116,zhang2025enhancinginformationretrievalquantumoptical}
\begin{equation}\label{eq:Fgd}
F(t;\varphi,\omega,\epsilon,\eta) = 2\eta\kappa\int_0^t d\tau\,\mathbb{E}\left[(\partial_\omega\mathbf{r}^\intercal)\mathbf{B}(\partial_\omega\mathbf{r})\right]. 
\end{equation}
This formula predicts that the CFI grows linearly at long times 
\begin{equation}
F(t;\varphi,\omega,\epsilon,\eta)  \sim k_F(\varphi,\omega,\epsilon,\eta)  t, 
\end{equation}
where the corresponding growth rate 
\begin{equation}
k_F = 2\eta\kappa\lim_{t\to+\infty} \mathbb{E}\left[(\partial_\omega\mathbf{r}^\intercal)\mathbf{B}(\partial_\omega\mathbf{r})\right]
\end{equation}
is a periodic function  of $\varphi$ with period $\pi$, i.e., $k_F(\varphi+\pi,\omega,\epsilon,\eta) = k_F(\varphi,\omega,\epsilon,\eta)$. In Fig.~\ref{fig:dist}(b) we plot $k_F$ versus $\varphi$, marking the two specific phases from panel (a) with a blue dot ($\varphi=0.1072$) and a red square ($\varphi=1$), respectively. Importantly, $k_F$ exhibits a characteristic double-peak structure, reaching its maximum at $\varphi=0.1072$ while nearly vanishing at $\varphi=1$. We emphasize that this characteristic profile, and consequently the existence of an optimal homodyne phase 
\begin{equation}\label{eq:vphiopt}
\varphi_\text{opt}(\omega,\epsilon,\eta) = \text{argmax}_\varphi k_F(\varphi,\omega,\epsilon,\eta),
\end{equation}
around which $k_F$ is significantly amplified, can be unequivocally attributed to the emergence of special critical points that effectively evade the measurement backaction, termed backaction-evading critical points in our previous work~\cite{zhang2025enhancinginformationretrievalquantumoptical}. 

The results in Fig.~\ref{fig:dist} therefore establish a clear connection between the distribution of the EKF estimate and the long-time growth of the CFI: a larger growth rate $k_F$ leads to a more prominent peak in the distribution of $\tilde{\omega}_\text{ekf}(t)$ around the true frequency $\omega_\text{true}$. This connection suggests that for better sensing performance the homodyne phase should be set to $\varphi = \varphi_\text{opt}(\omega_\text{true},\epsilon,\eta)$, a strategy we shall adopt in the next section. This choice, however, can only be realized adaptively, as $\varphi_\text{opt}$ itself depends on the unknown frequency $\omega_\text{true}$. 

We also observe that the distribution for $\varphi = 0.1072$ exhibits a long-tail consisting of estimates that greatly exceed the true value. The inset in Fig.~\ref{fig:dist}(a) quantifies the significance of this tail, showing the percentage of estimates satisfying $\tilde{\omega}_\text{ekf}(t^*) > \omega_0$ among 2000 trajectories at $\varphi = 0.1072$ as a function of $\omega_0$. Approximately $37\%$ of the estimates exceed $\omega_0=1.3\kappa$. Currently, we are not able to provide a full theoretical explanation for the occurrence of this tail. Yet, its impact on the subsequent sensing protocol is minimal, because the protocol operates by numerically fitting the central peak of $P(\tilde{\omega}_\text{ekf}(t))$. On the other hand, the presence of this tail necessitates the use of a skewed distribution function when performing numerical fits, as we analyze in more detail in the following section.

\begin{figure}[t]
\includegraphics[clip,width=8.5cm]{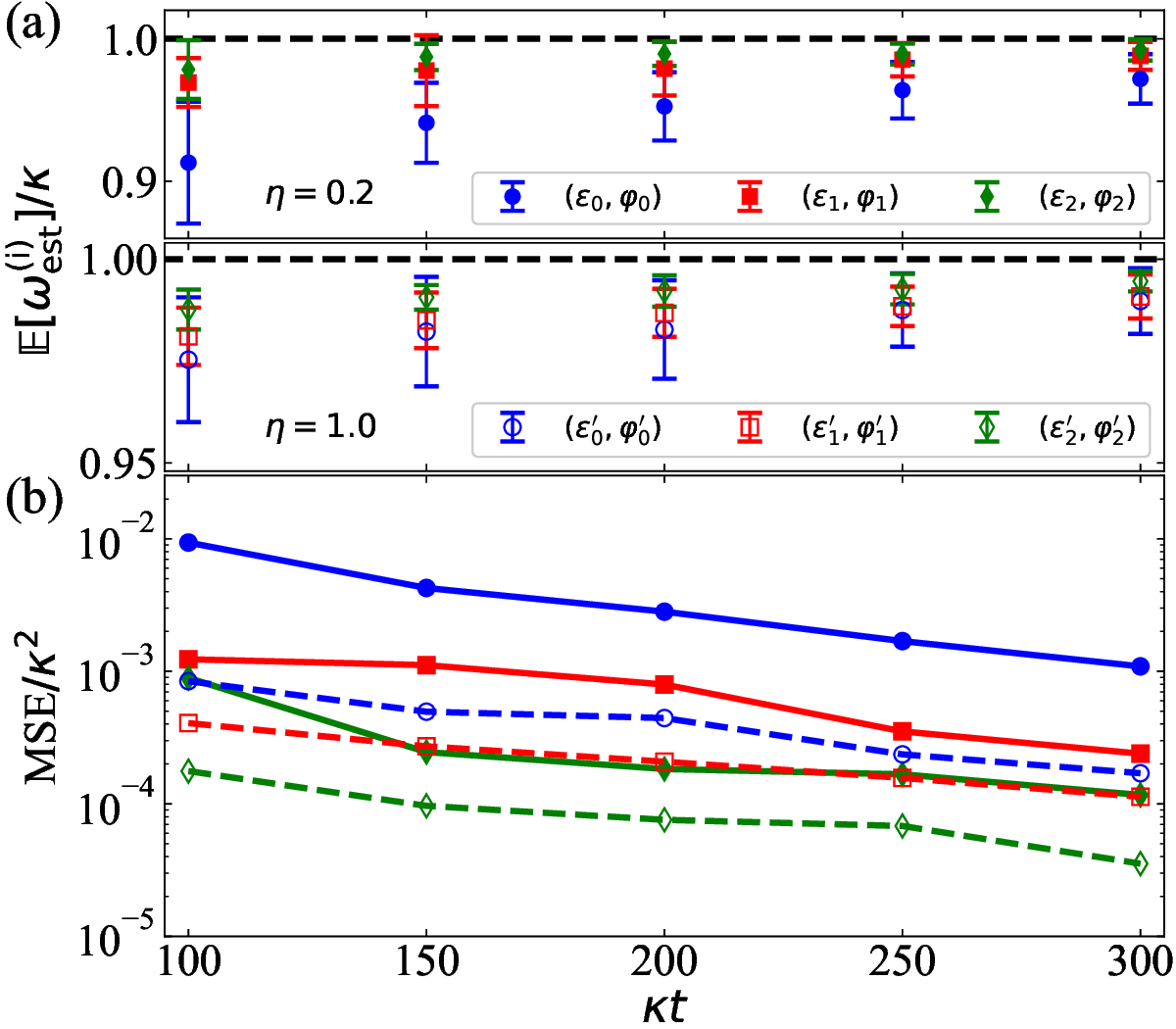}
\caption{(a) Mean $\mathbb{E}[\omega_\text{est}^{(i)}(t)]$ (symbols) and standard deviation $\text{Std}[\omega_\text{est}^{(i)}(t)]$ (error bars) and (b) the MSE (symbols) of the estimator $\omega_\text{est}^{(i)}$ obtained using the bootstrapping technique for both low ($\eta=0.2$, solid symbols, upper panel) and ideal ($\eta=1$, open symbols, lower panel) detection efficiencies. Here $N_\text{traj}=200$ for each run of the estimation. The control parameters $(\epsilon_i,\varphi_i)$ ($i\in\{0,1,2\}$) are identical to those used in Fig.~\ref{fig:adaptiveEKF}; the other parameter sets $(\epsilon_i',\varphi_i')$ are determined in Appendix~\ref{sec:sm_ideal}, with the following values: $(\epsilon_0',\varphi_0') = (0.760\kappa,2.996)$, $(\epsilon_1',\varphi_1') = (0.928\kappa,0.048)$, $(\epsilon_2',\varphi_2') = (1.021\kappa,0.139)$. 
}\label{fig:adaptiveEKFerror}
\end{figure}

\section{Global sensing protocol assisted by adaptive control}\label{sec:MCsampling}

We now present a global sensing protocol for criticality-enhanced estimation of the oscillator frequency. The protocol produces estimates from the distributions of $\tilde{\omega}_\text{ekf}(t)$ at each time instant $t$, assisted by adaptive control of the driving amplitude $\epsilon$ and the homodyne phase $\varphi$ to approach the unknown critical point for critical enhancement of sensing performance. The protocol operates iteratively, and the $i$th ($i\ge0$) iteration consists of the following three steps: 

\emph{Step 1.} Collect $N_\text{traj}$ trajectories of the EKF estimate, denoted as the set $\mathcal{S}_i = \{\tilde{\omega}_\text{ekf}^{(i,j)}(t) \vert j=1,\cdots,N_\text{traj}, t\ge0\}$. 

\emph{Step 2.} Compute the distributions $P(\tilde{\omega}_\text{ekf}^{(i,j)}(t))$ for all times $t\ge0$ using the dataset $\mathcal{S}_i$. We then fit these data using a fitting distribution function $P_\text{fit}(\omega)$. For each time $t$, the frequency estimate is defined as the frequency maximizing 
$P_\text{fit}(\omega)$, i.e., 
\begin{equation}\label{eq:mcekfomega}
\omega_\text{est}^{(i)}(t) = \text{argmax}_\omega P_\text{fit}(\omega). 
\end{equation}

\emph{Step 3.} Update the driving amplitude and the homodyne phase for the next iteration  according to the rules 
\begin{equation}\label{eq:selectrule}
\begin{array}{lll}
\epsilon_{i+1} &\displaystyle = \mathcal{E}\left(\epsilon_i, \epsilon_c\left(\omega_\text{est}^{(i)}(t_\text{large})\right)\right), \\
\varphi_{i+1} &\displaystyle = \varphi_\text{opt}\left(\omega_\text{est}^{(i)}(t_\text{large}), \epsilon_{i+1}, \eta\right), 
\end{array}
\end{equation}
where the update function $\mathcal{E}(\epsilon',\epsilon'')$ satisfies the property $\epsilon'< \mathcal{E}(\epsilon',\epsilon'') < \epsilon''$ whenever $\epsilon' < \epsilon''$, and we reiterate that the phase boundary is parametrized as $\epsilon_c(\omega) = \sqrt{\omega^2+\kappa^2/4}$~\cite{DiCandia2023,PhysRevLett.133.040801}. Here, $\omega_\text{est}^{(i)}(t_\text{large})$ represents the frequency estimation extracted in Step 2 at a sufficiently large time $t_\text{large}$, whose value should be determined case by case and typically requires $\kappa t_\text{large}\gg1$. This property of $\mathcal{E}(\epsilon',\epsilon'')$ ensures that the driving amplitude $\epsilon_{i+1}$ for the next iteration lies between the previous one $\epsilon_i$ and the latest guess $\epsilon_c(\omega_\text{est}^{(i)}(t_\text{large}))$ for the critical amplitude. A simple choice satisfying this property is $\mathcal{E}(\epsilon',\epsilon'') = (\epsilon'+\epsilon'')/2$ which is employed in the subsequent numerical simulations.

For the initial iteration ($i=0$), the driving amplitude $\epsilon_0$ is selected such that $\epsilon_0<\epsilon_c(\omega_l)$ to ensure the sensor operates in the normal phase. The  homodyne phase $\varphi_0$ is set to $\varphi_0 = \varphi_\text{opt}\left((\omega_l+\omega_h)/2, \epsilon_0, \eta\right)$ for the prior interval $I_\text{prior}$ defined in Eq.~(\ref{eq:prior}). 

In Fig.~\ref{fig:adaptiveEKF}, we illustrate the first three iterations ($i=0,1,2$) of the proposed protocol for the prior interval $I_\text{prior}=(0.7\kappa,2.3\kappa)$ and a low detection efficiency $\eta=0.2$. The true frequency is $\omega_\text{true}=\kappa$. Specifically, panels (a1)-(c1) show the time evolution of the distributions $P(\tilde{\omega}_\text{ekf}(t))$ till time $\kappa t^*=300$, while panels (a2)-(c2) display the corresponding distributions at the final time $t^*$, along with numerical fitting using a skew-normal distribution function defined as 
\begin{equation}\label{eq:skn}
P_\text{skn}(\omega) = A e^{-(\omega - \mu)^2/2\sigma^2} \left[1+\text{erf}\left(\alpha\frac{\omega-\mu}{\sigma}\right)\right], 
\end{equation}
whose amplitude, mean, deviation, and skewness degree are specified by the parameters $A$, $\mu$, $\sigma$, and $\alpha$, respectively. The error function $\text{erf}(\cdot)$ is introduced to account for potential asymmetry in the distributions. 

Note that, for all three iterations, we intensionally initialized the filter in the same manner as in Fig.~\ref{fig:singletrajs}; in particular, the initial guess $\tilde{\mathbf{x}}(0)$ was always set to $\tilde{\mathbf{x}}(0) = (0,0,1,1,0,(\omega_l+\omega_h)/2)^\intercal$. This choice facilitates the visualization of the process whereby the distribution $P(\tilde{\omega}_\text{ekf}(t))$ transitions from the wrong  initial guess $(\omega_l+\omega_h)/2$ to the true value $\omega_\text{true}$, as shown in Fig.~\ref{fig:adaptiveEKF}(a1)-(c1). In practical applications of the protocol, one could instead use the latest frequency estimate as the last entry of $\tilde{\mathbf{x}}(0)$ to improve sensing performance; for example, for the $1$st iteration in Fig.~\ref{fig:adaptiveEKF} one might set $\tilde{\mathbf{x}}(0) = (0,0,1,1,0,0.953\kappa)^\intercal$. 

We highlight that only $N_\text{traj}=200$ trajectories are used per iteration, confirming the practicality of the protocol under a moderate number (a few hundreds) of experimental repetitions. The initial driving amplitude is chosen as $\epsilon_0 = \epsilon_c(\omega_l=0.7\kappa)-0.1\kappa$; the subsequent two values $\epsilon_1=0.918\kappa$ and $\epsilon_2=1.009\kappa$ are determined via the selection function $\mathcal{E}(\epsilon',\epsilon'') = (\epsilon' + \epsilon'')/2$, using the frequency estimates $\omega_\text{est}^{(0)}(t^*)=0.953\kappa$ and $\omega_\text{est}^{(1)}(t^*)=0.980\kappa$ listed in panels (a2) and (b2) as its second arguments, respectively. As the frequency estimates become increasingly accurate, the sensor operates closer to the true critical point $\epsilon_c(\kappa)\approx1.118\kappa$. As a result, the distribution $P(\tilde{\omega}_\text{ekf}(t))$ develops a sharper peak around $\omega_\text{true}=\kappa$, providing a clear signature critical enhancement, i.e., the performance gain achieved by approaching the critical point. These results therefore validate the effectiveness of the proposed protocol, its robustness against detection inefficiency, and its capability to exploit critical enhancement in practice. 

A few remarks regarding the practical implementation of the proposed protocol are worth highlighting. First, the use of the skew-normal distribution $P_\text{skn}(\omega)$ is a biased choice, in the sense that it can only reproduce the main peak of $P(\tilde{\omega}_\text{ekf}(t))$ while failing to capture the long-tail on the right, as seen in the right panels in Fig.~\ref{fig:adaptiveEKF}. Although more sophisticated fitting functions could in principle be designed to reproduce the full statistics, in practice this brings little benefit, because the estimator defined by Eq.~(\ref{eq:mcekfomega}) relies primarily on the peak. In addition, the choice of the fitting function is not unique; any distribution function that captures the characteristics of the peak is suitable. Second, we highlight that the driving amplitude $\epsilon_0$ set during 
the initial iteration within the constraint $\epsilon_0 < \epsilon_c(\omega_l)$ is the only free control parameter in this protocol, while subsequent parameters (such as $\varphi_0$, $\varphi_1$, and $\epsilon_1$) should be determined adaptively.

We now further characterize the performance of the protocol using an empirical analysis. Specifically, we numerically examine the statistical properties of the estimator $\omega_\text{est}^{(i)}(t)$ in Eq.~(\ref{eq:mcekfomega}) in terms of its mean $\mathbb{E}[\omega_\text{est}^{(i)}(t)]$ and standard deviation $\text{Std}[\omega_\text{est}^{(i)}(t)]$, and quantify the estimation error in terms of the mean-squared error 
\begin{equation}\label{eq:mse}
\text{MSE} = \mathbb{E}[(\omega_\text{est}^{(i)}(t) - \omega_\text{true})^2]. 
\end{equation}
In particular we focus on how these quantities behave as the interrogation time increases, when detection efficiency is improved, and as the critical point is approached. 

In Fig.~\ref{fig:adaptiveEKFerror}(a) we present $\mathbb{E}[\omega_\text{est}^{(i)}(t)]$ (symbols) and $\text{Std}[\omega_\text{est}^{(i)}(t)]$ (error bars), while in Fig.~\ref{fig:adaptiveEKFerror}(b) we display the corresponding MSE as a function of time for both low ($\eta=0.2$) and ideal ($\eta=1$) detection efficiencies. All results are obtained by repeating the estimation procedure as in Fig.~\ref{fig:adaptiveEKF} for $50$ times, with the number of trajectories set to $N_\text{traj}=200$ and kept constant for each run. The control parameters $(\epsilon_i,\varphi_i)$ ($i\in\{0,1,2\}$) for $\eta=0.2$ are the same as those in Fig.~\ref{fig:adaptiveEKF}; the other parameter sets $(\epsilon_i',\varphi_i')$ for $\eta=1$ are presented in Appendix~\ref{sec:sm_ideal}. Although $50$ runs are insufficient to achieve convergence in the ensemble average $\mathbb{E}[\cdot]$, the results in Fig.~\ref{fig:adaptiveEKFerror} represent the empirical averages of these statistical quantities and correctly capture their dependence on the interrogation time, detection efficiency, and proximity to the critical point.

In all cases considered, we observe that with increasing interrogation time, improved detection efficiency, or as the critical point is approached, the estimator $\omega_\text{est}^{(i)}(t)$ in Eq.~(\ref{eq:mcekfomega}) becomes simultaneously less biased and more precise. This is evident as $\mathbb{E}[\omega_\text{est}^{(i)}(t)]$ converges towards the true value $\omega_\text{true}=\kappa$, while $\text{Std}[\omega_\text{est}^{(i)}(t)]$ tends to zero. Consequently, the corresponding MSE also decreases significantly under these favorable conditions.

\section{Conclusions}\label{sec:conclusions}

We have investigated global frequency sensing using a monitored KPO as the sensor and an EKF to produce the frequency estimate. Our analysis of the temporal convergence of this EKF reveals that, when used in isolation, it does not enable reliable single-shot estimation. However, an examination of its statistical properties highlights the potential for accurate estimation by leveraging the distribution of the EKF outputs. By complementing this technique with an adaptive control strategy for the KPO parameters, we have developed a protocol that achieves criticality-enhanced frequency estimation. Our numerical simulations confirm that the protocol yields accurate estimates even with a moderate number of experimental repetitions and under low detection efficiency, and further demonstrate its ability to harness critical enhancement, i.e., its performance improves as the critical point is approached.

Future work can further explore single-shot estimation by integrating the EKF with real-time feedback control, a strategy already successfully applied to atomic magnetometers~\cite{PhysRevA.102.063716,amorosbinefa2025trackingtimevaryingsignalsquantumenhanced}. It would also be interesting to extend the analysis beyond the Gaussian regime by accounting for corrections induced by finite Kerr nonlinearity.

\begin{acknowledgments}
C.Z. acknowledges support from CPSF (Grants No. 2025M1773419). M.C. acknowledges support from NSFC (Grants No. 11935012 and No. 12088101) and NSAF (Grant No. U2330401). 
\end{acknowledgments}

\appendix
\begin{figure*}%[t!]
\includegraphics[clip,width=0.9\textwidth]{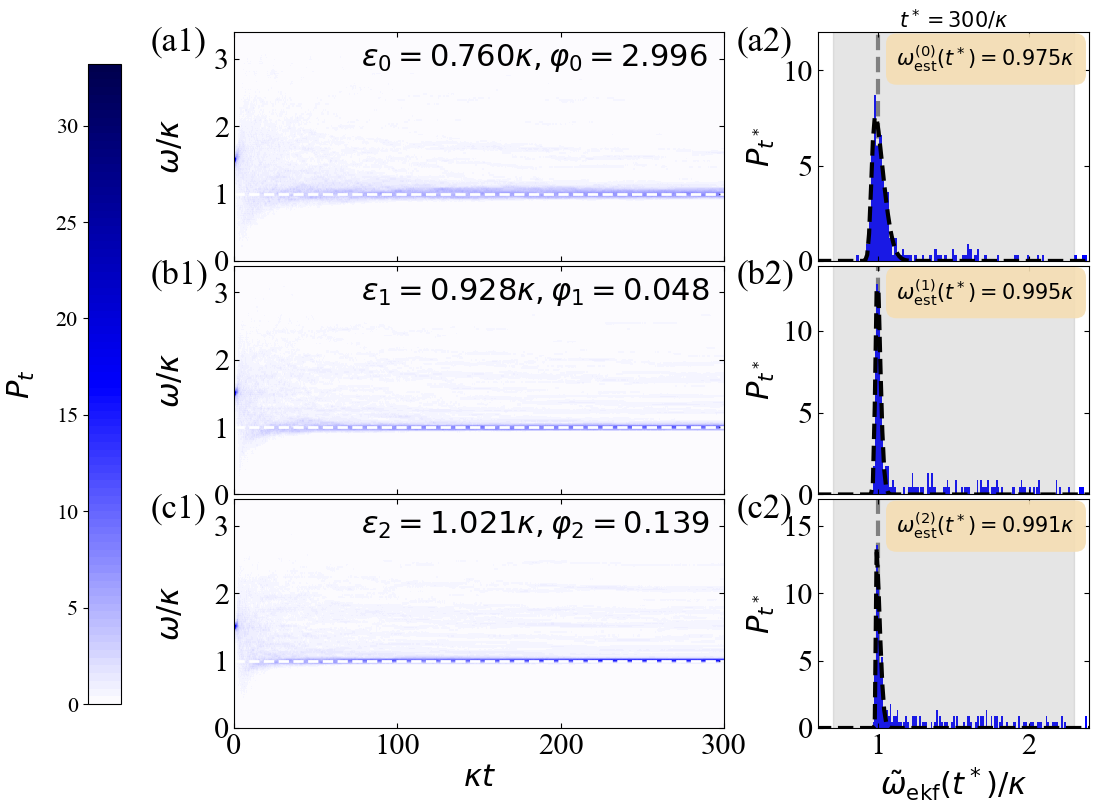}
\caption{(a1)-(c1) Time evolution of the distribution $P_t\equiv P(\tilde{\omega}_\text{ekf}(t))$ for the first three iterations ($i=0,1,2$) of the proposed protocol, obtained from $N_\text{traj}=500$ trajectories of $\tilde{\omega}_\text{ekf}(t)$. Here, the detection efficiency is $\eta=1$ and the prior interval is $I_\text{prior} = (0.7\kappa, 2.3\kappa)$. The control parameters $(\epsilon_i', \varphi_i')$ are determined according to the rules~(\ref{eq:selectrule}), using the update function $\mathcal{E}(\epsilon',\epsilon'') = (\epsilon' + \epsilon'')/2$ and the initial driving amplitude $\epsilon_0' = \epsilon_c(\omega_l=0.7\kappa)-0.1\kappa$. (a2)-(c2) Corresponding distributions at time $t^*=300/\kappa$ with black dashed lines showing numerical fits to the skew-normal distribution in Eq.~(\ref{eq:skn}). Other simulation parameters are the same as those in Fig.~\ref{fig:adaptiveEKF}. 
}\label{fig:adaptiveEKF_ideal}
\end{figure*}
\section{Explicit expressions of the vectors in Eq.~(\ref{eq:stateeq}) and the Jacobian matrix in Eq.~(\ref{eq:ekfeqs})}\label{sec:sm_cm}

The vectors $\mathbf{F}_\mathbf{x}$, $\mathbf{G}_\mathbf{x}$ and $\mathbf{H}$ in Eq.~(\ref{eq:stateeq}) are defined as
\begin{equation}\label{eq:Fx}
\mathbf{F}_\mathbf{x} = \begin{pmatrix}
-\kappa X/2 + (\omega-\epsilon)P \\
-(\omega+\epsilon)X - \kappa P/2 \\
2(\omega-\epsilon)\sigma_{xp} - \kappa(\sigma_x-1) - \eta\kappa \sigma_1^2 \\
-2(\omega+\epsilon)\sigma_{xp} - \kappa(\sigma_p-1) - \eta\kappa \sigma_2^2 \\
-(\omega+\epsilon)\sigma_x + (\omega-\epsilon)\Sigma_p - \kappa\sigma_{xp} - \eta\kappa\sigma_1\sigma_2 \\
0
\end{pmatrix}, 
\end{equation}
\begin{equation}
\mathbf{G}_\mathbf{x} = \left(\sqrt{\eta\kappa/2}\sigma_1,\sqrt{\eta\kappa/2}\sigma_2,0,0,0,0\right)^\intercal, 
\end{equation}
and 
\begin{equation}
\mathbf{H} = \left(\sqrt{2\kappa\eta}\cos\varphi, \sqrt{2\kappa\eta}\sin\varphi, 0, 0, 0, 0\right), 
\end{equation}
where we defined $\sigma_1 = \cos\varphi(\sigma_x-1) + \sin\varphi\sigma_{xp}$ and $\sigma_2 = \sin\varphi(\sigma_p-1) + \cos\varphi\sigma_{xp}$. 

In addition, the explicit form of the Jacobian matrix $\nabla\mathbf{F}_{\mathbf{x}}$ in Eq.~(\ref{eq:ekfeqs}) is
\begin{widetext}
\begin{equation}
% \mathbf{F}'_{\mathbf{x}}
\nabla\mathbf{F}_{\mathbf{x}} = \begin{pmatrix}
-\kappa/2 & \omega-\epsilon & 0 & 0 & 0 & P \\
-(\omega+\epsilon) & -\kappa/2 & 0 & 0 & 0 & -X \\
0 & 0 & -\kappa - 2\kappa\eta\cos\varphi\sigma_1 & 0 & 2(\omega-\epsilon)-2\kappa\eta\sin\varphi\sigma_1 & 2\sigma_{xp} \\ 
0 & 0 & 0 & -\kappa - 2\kappa\eta\sin\varphi\sigma_2 & -2(\omega+\epsilon)-2\kappa\eta\cos\varphi\sigma_2 & -2\sigma_{xp} \\ 
0 & 0 & -(\omega+\epsilon)-\kappa\eta\cos\varphi\sigma_2 & (\omega-\epsilon) - \kappa\eta\sin\varphi\sigma_1 & -\kappa-\kappa\eta\cos\varphi\sigma_1-\kappa\eta\sin\varphi\sigma_2 & -\sigma_x+\sigma_p \\
0 & 0 & 0 & 0 & 0 & 0 
\end{pmatrix}. 
\end{equation}
\end{widetext}

\section{Numerical results for ideal detection}\label{sec:sm_ideal}

Panels (a1-c1) in Fig.~\ref{fig:adaptiveEKF_ideal} show the time evolution of $P(\tilde{\omega}_\text{ekf}(t))$ for ideal detection efficiency $\eta=1$, while panels (a2-c2) display the corresponding distributions at time $t^*=100/\kappa$, along with their numerical fits to the skew-normal distribution in Eq.~(\ref{eq:skn}). The control parameters $(\epsilon_i',\varphi_i')$ ($i\in\{0,1,2\}$) are determined according to the rules~(\ref{eq:selectrule}), using the update function $\mathcal{E}(\epsilon',\epsilon'') = (\epsilon' + \epsilon'')/2$ and the initial driving amplitude $\epsilon_0' = \epsilon_c(\omega_l=0.7\kappa)-0.1\kappa$.

\bibliography{refs}

\end{document}